# The Structural Complexity of $(Na_{0.5}Bi_{0.5})TiO_3$-$BaTiO_3$ as Revealed by Raman Spectroscopy

Ben Wylie-van Eerd[1], Dragan Damjanovic[2], Naama Klein[2], Nava Setter[2] and Joe Trodahl[1,2]

[1]MacDiarmid Institute of Advanced Materials and Nanotechnology, Victoria University, PO Box 600, Wellington, New Zealand

[2]Ceramics Laboratory, Swiss Federal Institute of Technology – EPFL, 1015 Lausanne, Switzerland

Abstract

The structural phase diagram of the Pb-free ferroelectric $(Na_{½}Bi_{½})_{1-x}Ba_xTiO_3$ (NBT-BT), x<0.1, has been explored by Raman spectroscopy at temperatures from 10 to 470 K. The data provide clear evidence for a proposed temperature-independent morphotropic phase boundary at x ≈ 0.055. However, there is no evidence for a structural phase transition across T ≈ 370 K for x > 0.055, where bulk-property anomalies appear to signal a transition to a nonpolar or antiferroelectric phase. The results identify that the phase above 370 K shows short-range ionic displacements that are identical to those in the long-range-ordered phase below 370 K. These conclusions provide a natural interpretation of the weak piezoelectric response in this system and have important implications for the search for Pb-free piezoelectrics.

## I. Introduction

The lead-oxide based ferroelectric Pb(Zr,Ti)O$_3$ (PZT) is the archetypical example of a ferroelectric with a strong and only weakly temperature-sensitive piezoelectric response at temperatures of practical interest. Although it has been known and exploited for nearly fifty years the fundamental physics supporting its remarkable properties has been recently reassessed in support of a search for an environmentally-friendly Pb-free substitute. To date there is no known alternative that shows strong response over a wide temperature range demanded by the many applications of piezoelectric devices in acoustics, scanning microscopy, actuators and many more.[1]

The success of PZT is recognised to rely on a boundary separating rhombohedral (R) and tetragonal (T) phases,[2] with a monoclinic phase as a structural bridge between them.[3] In this region where the free energy difference between two structures is at a minimum[4-6] the piezoelectric response of the lattice is especially strong due to the easy polarization rotation between [001]$_c$ and [111]$_c$ pseudocubic directions.[7, 8] Further enhancement of the effective piezoelectric coefficients in multidomain materials typically used in most applications is due to domain-wall motion which is also strong near the R-T phase boundary.[2, 9] It is a bonus that this boundary depends only weakly on temperature so that there is no significant change in the piezoelectric coefficients from room temperature up to about 450 K.[2] Such a temperature-independent "morphotropic" phase boundary (MPB) is clearly beneficial in applications requiring operation across a large temperature range, so that the search for replacements has focussed on materials showing an R-T MPB with similar properties. This search has so far been unsuccessful. The most thoroughly studied lead-free system, (K,Na,Li)(Nb,Ta)O$_3$,[10] exhibits a boundary with a mixed morphotropic and polymorphic character, with properties that are unstable with temperature.[1] The parent system (K,Na)NbO$_3$ possesses a vertical MPB, but it separates two pseudo-orthorhombic phases[2, 11] and thus does not benefit from the polarization-rotation enhancement.

Recent *ab-initio* calculations have raised the question as to whether Pb may be an essential ingredient for MPB systems with large piezoelectric coefficients. The calculations suggest that in its ground state PbTiO$_3$ undergoes transitions from the tetragonal to a cubic phase via ferroelectric monoclinic and rhombohedral phases.[12, 13] It was proposed that the origin of the MPB in PbTiO$_3$ solid solutions such as PZT is in the tuning of these phase transitions to room temperature by internal chemical pressure applied by the modifying atoms. The implication of these results is that a PZT-like MPB may be exhibited by systems containing an end member with the same cubic-tetragonal phase-change complexity as PbTiO$_3$, further suggesting that Pb may in fact be especially effective in forming systems with differently-distorted crystal structures lying so close in energy. However, a strong piezoelectric response has been reported in (Ba,Ca)(Ti,Zr)O$_3$ (BCTZ) solid solution near a polymorphic boundary separating rhombohedral and tetragonal phases.[14] The authors explain the large piezoelectric effect in this material and in PZT-like systems by the termination of the phase boundary at the tricritical point where paraelectric cubic and ferroelectric rhombohedral and tetragonal phases meet. They argue that the BCTZ composition, which at room temperature lies in the vicinity of the tricritical point, possesses an exceptionally isotropic free energy profile facilitating polarization rotation between pseudocubic [001]$_c$ and [111]$_c$ axes.

In this paper we report Raman scattering studies on (Na$_{1/2}$Bi$_{1/2}$)$_{1-x}$Ba$_x$TiO$_3$ (NBT-BT), one of the most promising lead-free piezoelectrics. NBT is a relaxor material and its solution with ferroelectric BT promises large piezoelectric properties observed in other relaxor-ferroelectric solid solutions, such as Pb(Zn$_{1/3}$Mg$_{2/3}$)O$_3$-PbTiO$_3$ and Pb(Zn$_{1/3}$Mg$_{2/3}$)O$_3$-PbTiO$_3$.[15] The similarity of bismuth and lead ions, in terms of their large ionic radius, large number of electrons and a lone electron pair,[16, 17] are all believed to contribute to the distortion of the structure and thus to its piezoelectric response. The low-temperature (T < 470 K) phase diagram for this material has been reported as is shown in Figure 1a. A boundary between the ambient-temperature phases has been reported variously as between x=0.05 and 0.07, and since their rhombohedral and tetragonal assignments have been recently questioned[18-20] we identify them here by the labels I and II. Furthermore there are reports suggesting that the two phases are separated by a pseudocubic region.[21] The system is known to transform at high-temperatures (T$_C$ ≈ 600 K) to a nonferroelectric cubic phase through a number of intermediate phases.[20, 22-28] The exact nature of those phases, one of which was suggested to be antiferroelectric,[23, 25, 28] is still a matter of contention.[20] However, despite its apparent promise this solid solution has so far failed to display a large increase in the properties at the MPB region, for which the results presented here suggest structural reasons. We will show that in NBT-BT the x ≈ 0.06 boundary is a true, temperature-independent MPB in the practical temperature range, the first lead free material exhibiting such an MPB. However we find that the MPB does not terminate at a tricritical point involving a third, nonpolar phase; rather the Raman inference is that the structure above 370 K is a short-coherence-length version of the ambient-temperature phase for x ≥ 0.06. We find that the MPB displays the remarkable property that it is shifted by poling, rising above x = 0.06 in poled samples, reverting to x < 0.06 only after cycling above 370 K. Uncertainties of the phase-boundary composition and the reports of pseudocubic phases are likely to result from this irreversible structural change on cycling the temperature through 370 K.

## II. Experimental Details

Ceramic samples of NBT-BT with x=0, 2.5, 5, 6, 7 and 9% were prepared by standard solid state synthesis from Ba and Na carbonates and Ti and Bi oxides.[29] Data obtained from standard electrical measurements (polarisation hysteresis loops, piezoelectric and dielectric responses, described in detail in Ref. [29]) shown in Figure 2 are similar to those reported in the literature.[18, 22, 23, 30-32] The depolarization temperature at which permittivity exhibits a sharp jump as a function of temperature and above which samples develop pinched polarization-electric field hysteresis and poled samples lose macroscopic polarization is marked by $T_d$ in Figure 2. The phase diagram determined by those measurements is compared to that of Hiruma et al.[22] in Figs. 1a and 1b. The samples on which we report here include as-prepared unpoled pellets, pellets that were poled at 333 K with a field of 60 kV/cm applied for 5-10 minutes, and the same pellets after they have been depoled by cycling the temperature above 370 K.

Raman data were collected using 633 and 514.5 nm excitation with a Jobin-Yvon LabRam HR and a Jobin-Yvon T64000 spectrometers. For measurements above 80 K the pellets were placed in a model HMS 600 Linkam variable-temperature stage, and a closed-cycle cryocooler was used for lower temperatures. It is important to recognise that Raman spectroscopy is a probe of very local ionic configuration, in contrast with XRD. In the materials studied here, with both compositional and structural disorder broadening the features, we expect that Raman spectroscopy probes ionic configurations on length scales substantially shorter than 10 nm. Ionic displacements that are coherent over such a short length scale are easily missed in XRD investigations[22] so that Raman spectroscopy provides a useful complement to the larger-length-scale averaged structural configuration probed by XRD.

**III. Results and Discussion**

Figure 3a shows the ambient-temperature spectra from as-prepared (unpoled) pellets. We do not assign here the normal-mode vibrations that are responsible for the Raman lines, but note that these are already discussed in the literature.[33] Simple inspection of the spectra shows clearly altered patterns between 5 and 6% BT, signalled here by (1) a weakening and softening of the mode near 130 cm$^{-1}$, (2) a splitting of the mode near 240 cm$^{-1}$ into two lines, and (3) a moderate hardening of the modes near 600 cm$^{-1}$. These changes are seen even more clearly in the central frequencies in the inset of Fig. 3a, determined by fitting the spectra with Lorentzian lines. The mode frequencies change discontinuously across the transition, but there is no sign of any other structural changes, in direct contradiction with suggestions of a third pseudocubic phase; we will return to this point in connection with spectra from poled samples. Note that these spectra bear a very strong resemblance to those reported by Rout *et al.*,[33] which confirmed the existence of the phase boundary near 6% marked in Fig. 1c, but they were not extended to lower or higher temperatures.

Addressing now the temperature dependence of this phase boundary, we note that the spectra at 80 K shown in Figure 3b show the same behavior, with only the expected sharpening of the lines associated with reduced anharmonic effects. The data from 10 to almost 370 K have the same discontinuous shifts across x=5-6% as are seen in Figs. 3a and 3b. Thus the phase boundary in unpoled samples lies at x = 5.5 ± 0.5% from 10 to 350 K (filled circles in Fig. 1c), as good an example as exists of an ideal temperature-independent MPB. It is vertical within 0.5 % from 10 K to 350 K, with less polymorphic character than found in PZT.[34] Thus the MPB in NBT-BT shows immediately two of the properties to make it a promising candidate to replace PZT: (i) a vertical phase boundary (ii) separating phases of contrasting symmetry.

The spectral changes seen in the x = 5% pellet while raising the temperature from zone I are displayed in Figure 4a, showing a remarkable similarity to the spectral shifts across the 5-6% MPB between the zones I and II. Both the doubled 300 cm$^{-1}$ feature and the loss to lower frequency of the 130 cm$^{-1}$ line suggest that the Raman-active phase in zone III has exactly the local structure as does the ferroelectric phase II. Confirmation is found in the complete absence of discontinuous spectral changes across the zone II-zone III boundary, see Figure 4b. The frequencies, and remarkably even the scattering amplitudes, do not vary with temperature across this boundary that has been proposed from dielectric/piezoelectric studies; the features show only a gradual broadening as the temperature rises. There is no signature at all of changes of ion-site symmetries expected when crossing a ferroelectric/antiferroelectric boundary.[35, 36] The image that emerges is of a local structure that is unchanged across the zone II-zone III boundary, and that the transition seen in data from electrical-properties measurements is associated with ionic displacements that lose their long-range coherence across the boundary.

It is of interest to consider the scale over which there might be the zone-II-like distortion in order for the zone III to retain the zone-II Raman spectrum. We propose that in this system there is a strong propensity to form frequent domain boundaries within each crystallite, forming domains of typical size *d*; this then takes the role of a structural coherence length. The Raman spectra following from such a structure is expected to show broadening of the zone-center signals, as modes within $\pi/d$ of the zone center are rendered Raman active by the loss of full translational periodicity. Estimating the broadening effect from this mechanism requires the dispersion of the Raman active phonon branches, which is not available. However, reasonable estimates of the optic branch band widths (e.g. 25% of the zone-center frequency) and recognition that the dispersion is quadratic near the zone center establishes that domains as small as a few nm would broaden the Raman features by only about 10 cm$^{-1}$. Such weak broadening would not be observed in the already very broad spectra of Figs. 3 and 4, leaving the Raman signal indistinguishable from the ambient-temperature phase II.

A suggested formation of domain boundaries should be consistent with macroscopic behaviour of NBT-BT on crossing zone II – zone III boundary, specifically with pinched polarization-electric field loops and macroscopic depolarization of the sample, Fig. 2, which were previously taken as manifestation of antiferroelectricity. Our Raman data do not support transition to an antiferroelectric phase. The simplest configuration of domain boundaries that can explain both the evolution of Raman spectra and pinched polarization loops / depolarization is 180° boundaries along with strong restoring force acting on the domains. This is consistent with reported observation of large reversible strain associated with pinched polarization loops above depolarization temperature in related system $(K_{0.5}Na_{0.5})NbO_3$-modified NBT-BT, which is interpreted by motion of polar regions in a nonpolar matrix.[37] Note that the zone II-zone III boundary is different from the one reported in other relaxor materials (e.g., $Pb(Sc_{0.5}Nb_{0.5})O_3$, Ref. [38]). In those materials dielectric anomaly and pinched loops similar to those exhibited by NBT-BT are accompanied by a clear first order phase transition between ferroelectric and nonpolar phase, which is not the case for II - III boundary in NBT-BT.

The boundary between zones II and III (Figs. 1a and 1b) has been previously located primarily by dielectric, ferroelectric and piezoelectric measurements[22] and only on poled samples by heating from ambient temperature: examples of such measurements are shown in Fig. 2 for the samples used in this study. Poled samples are depoled by cycling to above 370-450 K after which they no longer show any clear signature of that II-III boundary. To see whether the boundary is detected by Raman spectroscopy in poled samples, we have repeated the measurements on pellets that have been poled at 333 K. The resulting spectrum for the poled 6% sample is compared to unpoled 5 and 6% spectra in Figure 5a. The 6% poled spectrum bears a remarkable resemblance to the spectra in depoled samples with 5% BT and other zone I samples; it is immediately clear that poling has shifted the phase boundary to the right. On closer inspection the poled 6% BT sample shows slightly broader spectral lines than is seen in the x = 0.05 sample, indicating that the 6% BT poled sample is mixed-phase, that there is still some fraction of material with the zone II structure. The mixed-phase nature is signalled most clearly by the shoulders on both sides of the 280 cm$^{-1}$ feature. The fit of that line in the 80 K spectrum to a weighted sum of zone I and zone II spectra, shown in Fig. 5b, yields an approximately 20% admixture of the phase II spectrum.

Upon cycling the temperature above 375 K, which depoles the material, the I-II MPB lies again between 5 and 6% BT. In contrast there is no sign of this behaviour at 7% BT; at that concentration the sample retains the local structure of phase II even after poling. It is important to note that both the zone I and zone II phases can support a remnant polarisation, and it is only the act of poling that alters the local structure of the x = 0.06 ceramic and takes it across the phase II-phase I boundary.

The structural instability found at x = 0.06 is partially mimicked at x = 0.05. At this composition the fully stable structure is that found also at lower BT concentrations, as is indicated by the uniform spectra across the low-concentration region of Fig 3. However, when returning to ambient temperature after cycling into zone III above 400 K the spectra show a mixed zone I-zone II character, returning to the pattern of Fig. 3 only after some days. That slow recovery is found not only in poled samples; it is identical also in both unpoled and depoled 5% BT samples. The two structurally distinct phases appear to have very similar free energies across the entire region $0.05 \leq x \leq 0.06$, a clear sign of an MPB.

It is interesting to compare and contrast the signatures of bistability we report here with XRD investigations of a 7% ceramic sample.[39] In that case an unpoled sample was reported to adopt a pseudocubic structure, but showed an electric-field induced transition to the tetragonal phase of zone II. Our Raman data show no difference between poled, unpoled or depoled spectra for this composition, signaling a continued zone II (tetragonal?) local structure for all three histories and highlighting the contrasting scales probed by Raman and XRD spectroscopies. More recently there has been a report of field-switchable structures in a lightly Mn-doped NBT-5.5%BT single crystal, which was switched from a zone I (rhomohedral?) structure to a zone II (tetragonal?) structure by application of a field along (001).[32] The MPB shift was then in the opposite sense to the present data, a result not unexpected since a field along (001) specifically favors a tetragonal phase. Finally we note that in some lead based materials poling shifts the MPB toward lower x, opposite to what is observed here in NBT-BT (e.g., $(1-x)Pb(Mg_{1/2}Nb_{1/2})O_3-xPbTiO_3$; Ref. [40]).

An important inference can be drawn from the phase diagram determined by the Raman measurements. It is now clear that the dielectric anomalies measured across the zone II-zone III boundary do not signal a phase transition, rather they follow from a loss of long-range coherence of the zone II structure. Thus this boundary does not involve a transition into a nonpolar phase, in direct contrast to either PZT or BCTZ, the two systems that have shown strong piezoelectric responses. The absence of a nonpolar phase has important consequences on the piezoelectric properties of NBT-BT. Firstly, it means that a tricritical (triple) point similar to the one in PZT and BCTZ formed by two polar and one nonpolar phase is lacking. In PZT and BCTZ free energy profile near the tricritical point is globally isotropic[14,41,42] leading to a large increase in shear and longitudinal piezoelectric coefficients.[42] Secondly, in these materials as well as in barium titanate, an end member of NBT-BT, the transition from tetragonal to cubic phase away from the tricritical point is accompanied by a large increase in longitudinal piezoelectric coefficient.[42-44] Neither of these enhancements happens close to the zone II - zone III boundary of NBT-BT because zone III is only macroscopically nonpolar while on the unit cell scale where properties enhancement takes place it is polar and with a short-range configuration identical to phase II. Thus the MPB does not terminate at a nonpolar phase, an observation signalling that the free energy function does not flatten globally as zone III is approached, with the result that the piezoelectric coefficients in NBT-BT are not particularly large.

## IV. Conclusions

We have reported a search for Raman evidence of structural changes in NBT-BT. We find a vertical MPB in unpoled samples to lie in the range of 5 and 6 % BT from 10 to 370 K, as vertical a phase boundary as exists in any system. The data show no structural development across the 0-9% BT range except for that discontinuity at the MPB; there is no signature at all of weakened structural displacements characteristic of the two phases near the MPB as has been suggested based on XRD. Furthermore no evidence exists in the data of a discontinuous structural phase-boundary change across the transition at 370 K for 6-9% BT as would be expected for the FE-AFE transition suggested on the basis of dielectric measurements. The data support instead a loss of long-range FE order and a rapid reduction of the correlation length of FE order parameter across that boundary. The phase above 370 K, which could be argued as nonpolar or antiferroelectric on the basis of bulk dielectric and piezoelectric response measurements, is in fact zone II polar phase at the short length scale probed by Raman scattering.

Ambient-temperature poling shifts the vertical MPB to lie in the range of 6-7%, signalling a close correspondence between the free energies of the two phases at 6% BT concentration as would be expected for an MPB system. A weaker instability is found also at 5% BT. The picture that emerges is of the MPB resulting from an instability between two polar phases, but without the full flattening of the free energy in three dimensions. We have related the relatively poor piezoelectric performance of this material to exactly that lack of full free energy flattening.

Acknowledgements

Financial support of the Swiss National Science Foundation, Swiss Priority Programme PNR62 (Project No. 406240-126091) and the European Network of Excellence "MIND" are gratefully acknowledged. The work in the MacDiarmid Institute is supported by the New Zealand Foundation for Research Science and Technology under the Centres of Research Excellence programme.

# References

*Joe.Trodahl@vuw.ac.nz

**Figure captions:**

Figure 1. The phase diagram of $(Na_{1/2}Bi_{1/2})TiO_3$-$BaTiO_3$. a) from dielectric and piezoelectric data, Ref. 22; b) from dielectric (triangles) and piezoelectric (squares) data taken on samples used in this work, Ref. 29; c) from the present Raman results. Raman data are shown as circles, filled along the MPB. Note that there were data taken every 20 K along the MPB, and for clarity only very few are shown here. The MPB positions for unpoled and poled samples are shown as solid and dashed lines, respectively.

Figure 2 (color online). Dielectric permittivity for (a) poled and (b) unpoled NBT-6% BT samples. $T_d$ marks the depolarization temperature. Polarization – electric field hysteresis loops are shown below and above depolarization temperature for (c) NBT-5% BT and (d) NBT – 7% BT. Pinched loops are observed above $T_d$. (e) Illustration of the macroscopic depolarization of the sample as observed in poled NBT- 7%BT sample. The figure shows the height of the sample's conductance, $\Delta G_{max}$, measured at the series resonance frequency of the planar mode vibrations of a poled disk as a function of temperature. The resonance becomes very small (but not zero) at the depolarization temperature $T_d$ indicating strong reduction of the macroscopic polarization.

Figure 3. a) Ambient-temperature Raman spectra for $(Bi_{1/2}Na_{1/2})_{1-x}Ba_xTiO_3$, $0 \leq x \leq 9\%$. The inset shows the frequency shifts, as determined by a multiline fits to the spectra. b) 80 K data.

Figure 4. a) Trends with temperature of the spectra from the 5% BT sample, and (inset) the fitted frequencies. Note the similarity to the results across the 5-6% boundary shown in Fig. 3. b) Spectra across the putative Zone II-III boundary at 350 K for x = 6% (see Fig. 1). The data show only the expected gradual anharmonic broadening with rising temperature; there are no discontinuities near 350 K.

Figure 5 (color online). a) 80 K Raman spectrum of a poled 6% sample compared with zone I (5% depoled) and zone II (6% depoled) spectra. The 6% BT poled spectrum resembles the zone I spectrum much more closely than it does the 6% BT spectrum. b) the poled spectrum of Fig. 5a is well reproduced by an admixture of 80% zone I and 20% zone II spectra of Fig 5a. The fit was chosen to reproduce the peak centred at 250 cm$^{-1}$, and is not fully perfect for the peaks centred on 550 cm$^{-1}$.

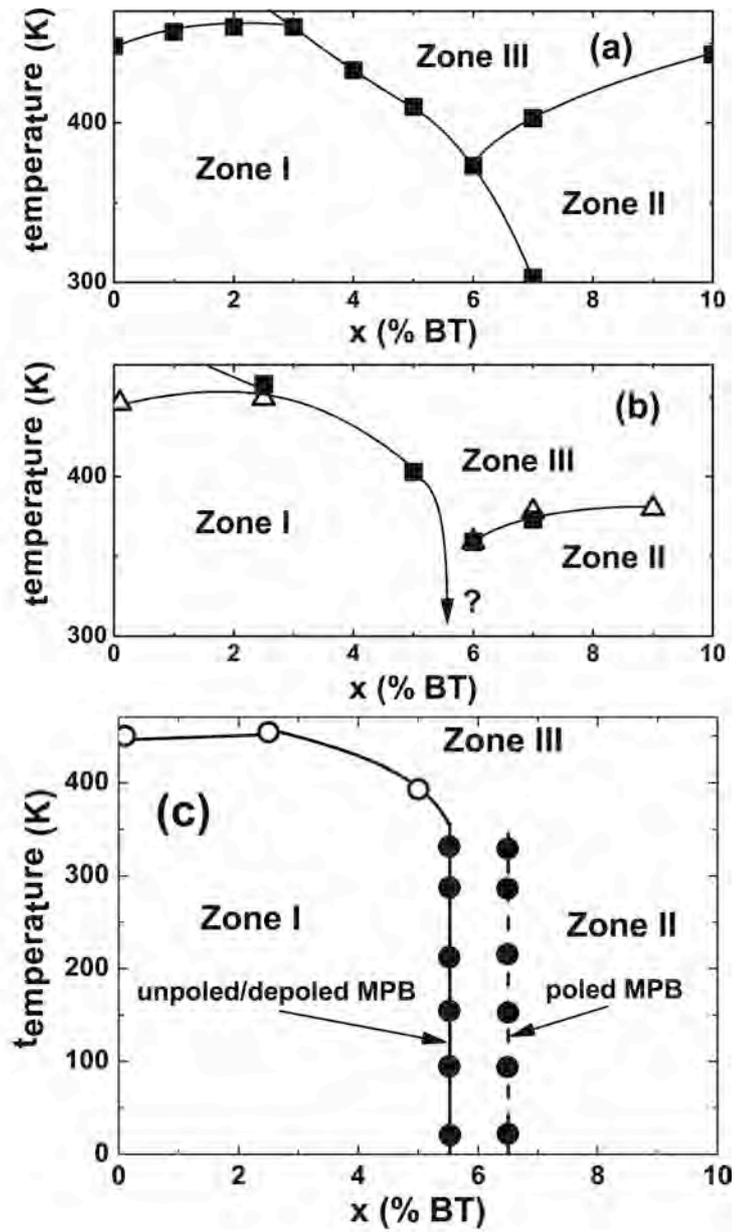

Figure 1.

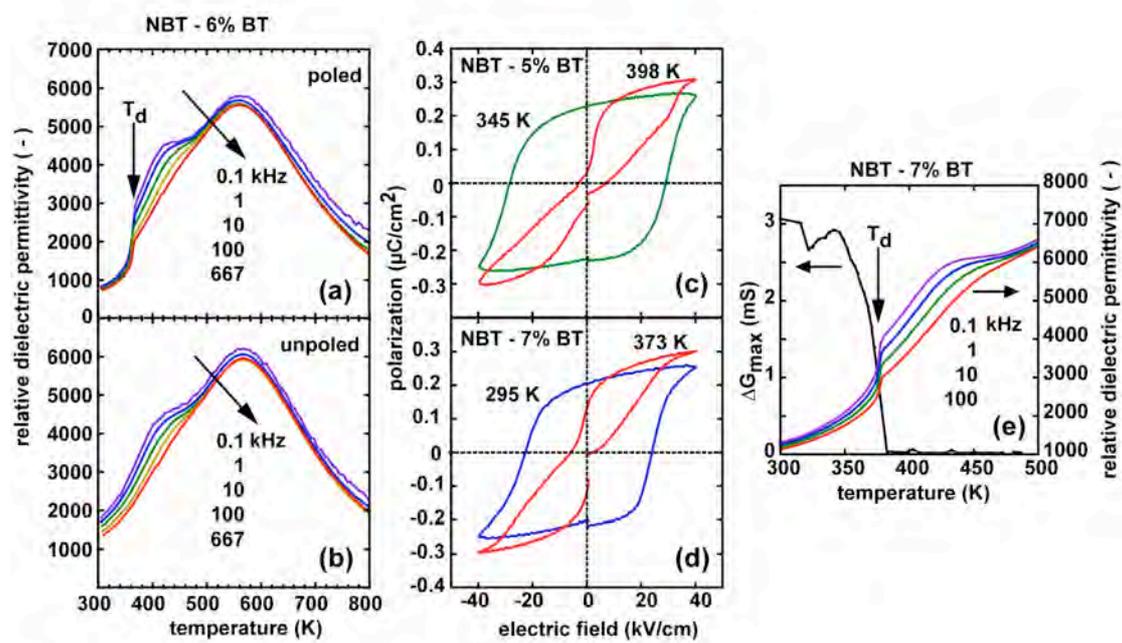

Figure 2.

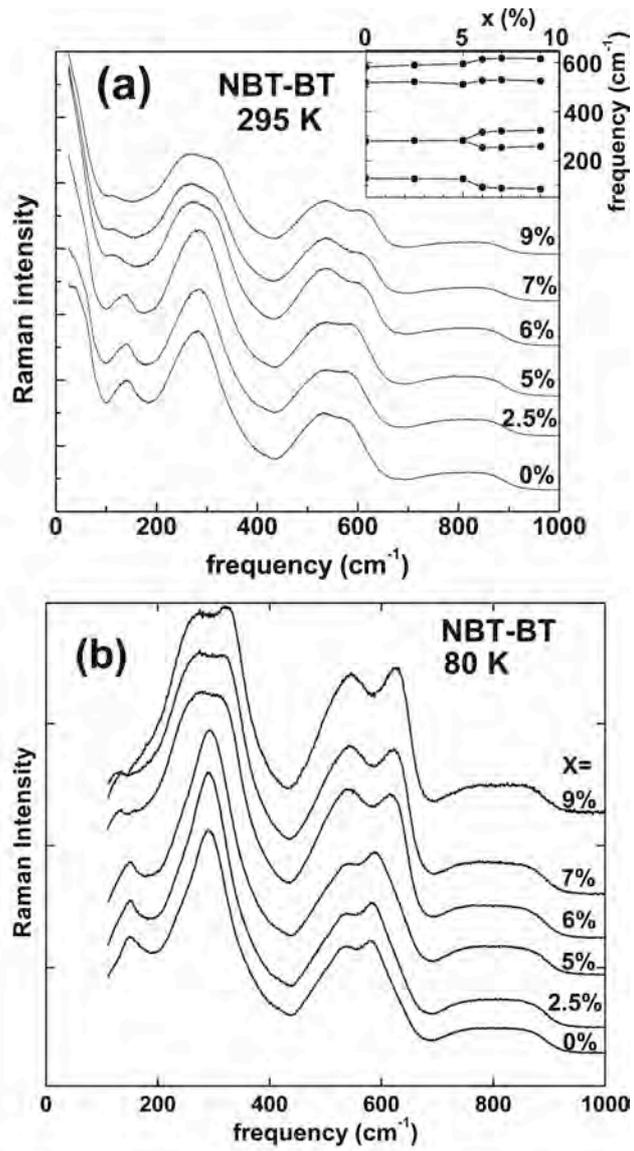

Figure 3.

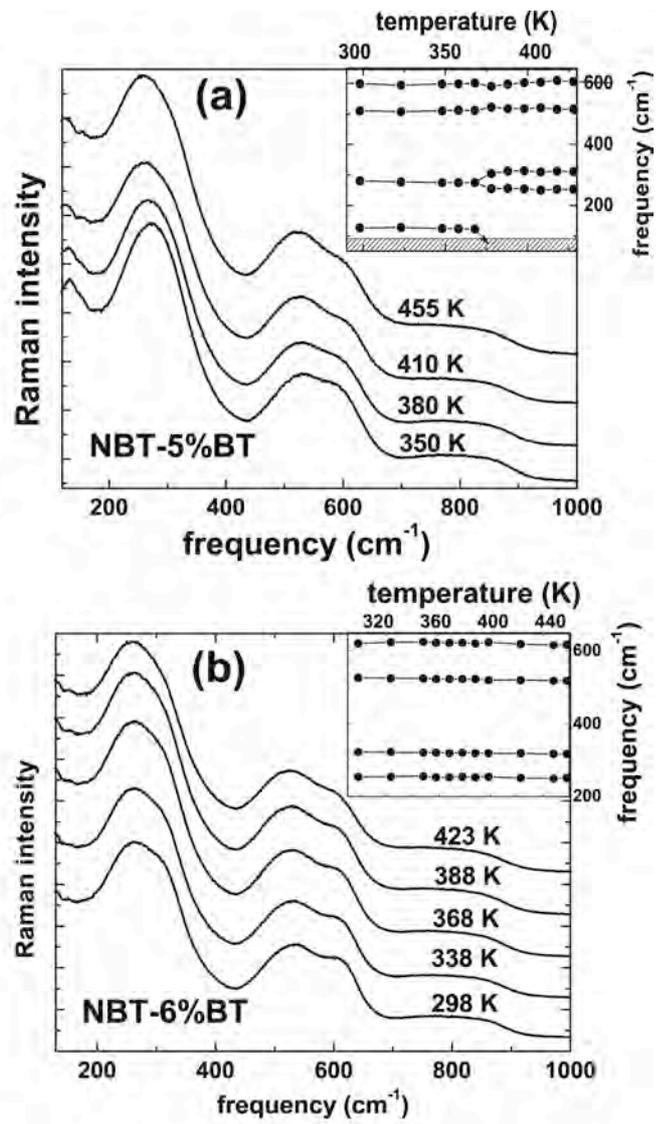

Figure 4

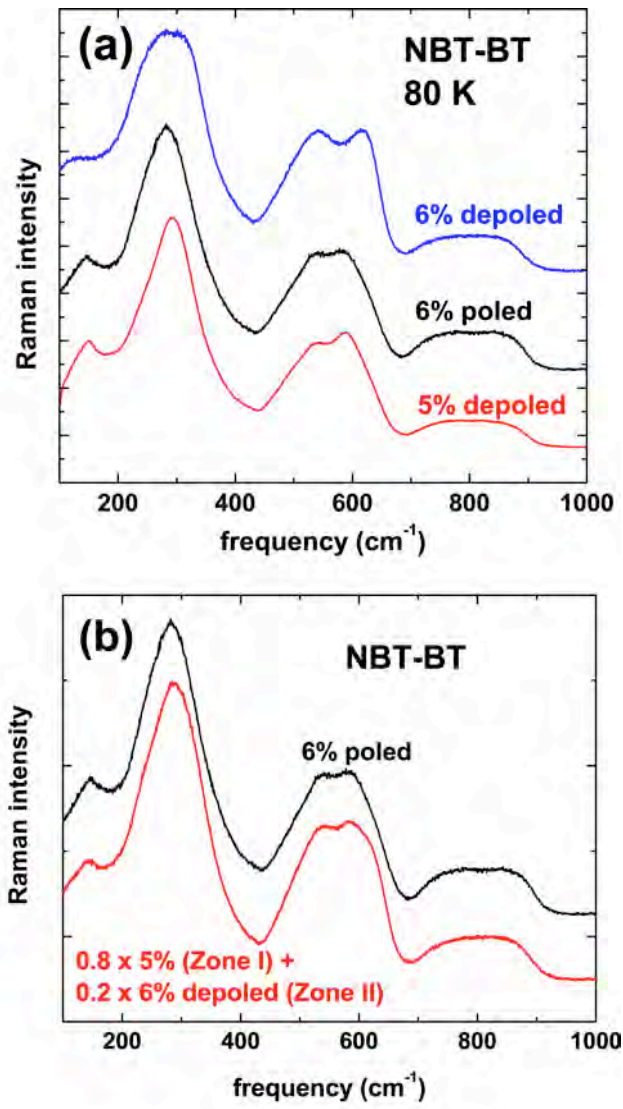

Figure 5